\documentclass[review]{elsarticle}

\usepackage{lineno,hyperref}
\modulolinenumbers[5]

\journal{Nuclear Physics B}

\usepackage{amssymb}

\begin{document}

\begin{frontmatter}

\title{Hierarchy of the low-lying excitations
for the $(2+1)$-dimensional $q=3$ Potts model
in the ordered phase
}


\author{Yoshihiro Nishiyama}

\address{Department of Physics, Faculty of Science,
Okayama University, Okayama 700-8530, Japan}



\begin{abstract}
The 
$(2+1)$-dimensional
 $q=3$ Potts model
was simulated with the exact diagonalization method.
In the ordered phase,
the elementary excitations (magnons) are attractive,
forming a series of bound states in the 
low-energy spectrum. 
We investigate the low-lying spectrum
through a dynamical susceptibility,
which is readily tractable with the exact diagonalization method
via the continued-fraction expansion.
As a result,
we estimate the series of (scaled) mass gaps,
$m_{2,3,4}/m_1$ 
($m_1$: single-magnon mass), in proximity to the transition point.
\end{abstract}

\begin{keyword}

75.10.Jm 	
75.40.Mg 
05.50.+q , 
05.70.Jk 
\end{keyword}

\end{frontmatter}


\section{\label{section1}Introduction}

In the ordered phase for an Ising ferromagnet,
the elementary excitations
(magnons) are attractive,
forming a series of
bound states
(composite particles)
with the mass gaps,
$m_1<m_2<\dots$ ($m_1$: single-magnon mass).
In fact,
for the $(1+1)$-dimensional Ising model
\cite{Zamolodchikov98,Delfino04,Fonseca03},
there appear eight types of excitations
with the (scaled) mass gaps
\begin{eqnarray}
m_2/m_1&=&2\cos\pi/5 \\
m_3/m_1&=&2\cos\pi/30 \\
m_4/m_2&=&2\cos7\pi/30 \\
m_5/m_2&=&2\cos2\pi/15 \\
m_6/m_2&=&2\cos\pi/30 \\
m_7/m_2&=&4\cos\pi/5\cos7\pi/30 \\
m_8/m_2&=&4\cos\pi/5\cos2\pi/15 
            ,
\end{eqnarray}
under a properly scaled magnetic field
so as to preserve the integrability;
namely,
the scaled gap ratio,
$m_{2,3,\dots}/m_1$,
displays a universal hierarchical character.
For a quasi-one-dimensional
ferromagnet, CoNb$_2$O$_6$,
the primary one, $m_2/m_1=1.61\dots$ (golden ratio),
was confirmed by means of the
inelastic neutron scattering \cite{Coldea10}.
For the $(2+1)$-dimensional counterpart,
such rigorous information is not available,
and various approaches have been made 
so as to fix the hierarchical
structure $m_{2,3,\dots}/m_1$
\cite{
Caselle99,Lee01,Caselle02,Nishiyama08,
Dusuel10,Nishiyama14,Rose16,Nishiyama16}.
Meanwhile,
it turned out that the spectrum
for
the three-state ($q=3$) Potts model 
exhibits a hierarchical character as well
\cite{
Chim92,Delfino08,Lepori09,
Rutkevich15,Lencses15,Falcone07}.
Related results are recalled afterwards.

According to the Monte Carlo simulations 
in $(2+1)$ dimensions \cite{Falcone07,Falcone07b},
the hierarchy $m_{1,2,\dots}$
of the Potts model
and
that of the pure gauge theory
look alike.
Actually, for the Z$_2$ case in $(2+1)$ dimensions,
a duality relation \cite{Wegner71,Balian75,Fradkin78} 
does hold, ensuring
the correspondence between them.
Generically \cite{Svetitsky82},
the SU$(N)$ gauge theory displays a global Z$_N$ symmetry 
(center of SU$(N)$), which immediately
establishes  a relationship between them;
for $N \ge 3$, the transition would not be critical,
and the universality idea does not apply nonetheless.
Meanwhile,
an extensive lattice-gauge-theory simulation reveals
the ``weak $N$ dependence'' \cite{Athenodorou15} of the SU$(N)$ theory,
suggesting a robustness of the hierarchy, $m_{2,3,\dots}/m_1$.
On the one hand,
 the $q=3$ Potts model exhibits 
an ``approximate universality'' \cite{Hamer92},
even
though the phase transition is  of first order.
Hence,
it is expected that the hierarchy
should display a model-independent character to some extent.

\begin{figure}
\includegraphics[width=100mm]{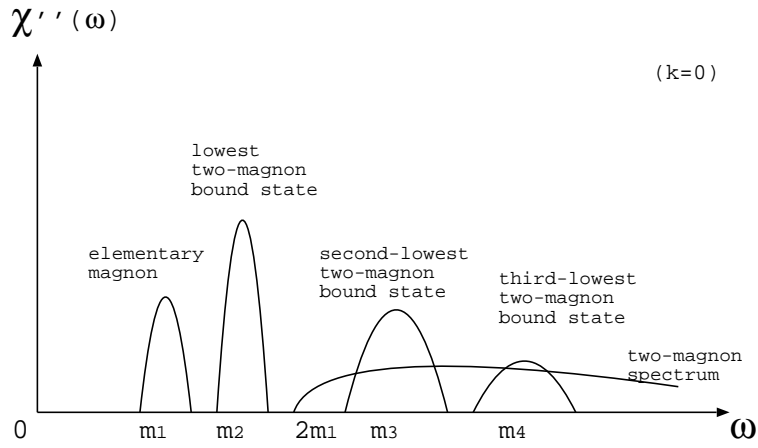}%
\caption{
\label{figure1}
A schematic drawing of a dynamical susceptibility
for the Potts model (\ref{Hamiltonian})
in the ordered phase
within the zero-momentum ($k=0$) sector
is presented.
There appear hierarchical peaks with the mass gaps
$m_{1,2,\dots}$:
The elementary excitation $m_1$
corresponds to the single magnon,
which 
forms a series of bound states $m_{2,3,\dots}$.
The two-magnon spectrum extends above 
$\omega>2m_1$.
The excitations $m_{1,2,\dots}$
may have a relevance to
the glueball spectrum
for the (pure) gauge field theory
\cite{Falcone07,Falcone07b}.
}
\end{figure}

In this paper,
we investigate the
$(2+1)$-dimensional $q=3$ Potts model
\cite{Hamer90,Dai14}
by means of 
the numerical diagonalization method.
The method allows us to
evaluate
the
dynamical susceptibilities, 
Eqs. (\ref{dynamical_susceptibility_1})
and
(\ref{dynamical_susceptibility_2}),
 via the continued-fraction-expansion method
\cite{Gagliano87}.
In Fig. \ref{figure1},
we present a schematic drawing for a
spectral function.

To be specific,
we present the Hamiltonian
for the $(2+1)$-dimensional $q=3$ Potts model
\begin{equation}
\label{Hamiltonian}
{\cal H} = 
- \sum_{\langle ij \rangle} 
         \frac{2}{3}
	 \cos \left(\frac{2\pi}{3}(L_i-L_j)\right)
          - \lambda \sum_{i=1}^N (R_i^+ +R_i^-)
           -H 
\sum_{i=1}^N \frac{2}{3} \cos
  \frac{2\pi L_i}{3}  
.
\end{equation}
Here, the operator 
\begin{equation}
L_i=
\left(
\begin{array}{ccc}
0 & 0 & 0 \\
0 & 1 & 0 \\
0 & 0 & 2 
\end{array}
\right)
,
\end{equation}
is placed at each square-lattice point $i=1,2,\dots,N$;
namely, 
the base $|l_i\rangle$ ($l_i=0,1,2$)
satisfies
$L_i |l_i\rangle =l_i |l_i\rangle$. 
The summation
$\sum_{\langle ij\rangle}$
runs over all possible nearest neighbor pairs $\langle ij\rangle$.
The operator
$R_i^\pm$ induces the transition
$R_i^\pm |l_i\rangle = | l_i \pm 1 \ {\rm mod} \  3 \rangle $,
and the parameter $\lambda$ denotes the corresponding 
coupling constant.
This model exhibits the first-order phase transition
at $\lambda = \lambda_c= 0.8758(14)$ ($H=0$) \cite{Hamer92},
which separates the 
ordered ($\lambda<\lambda_c$) and 
disordered ($\lambda>\lambda_c$) phases.
An infinitesimal magnetic field $H=20/L^{2.5}$
($L$: linear dimension of the finite-size cluster)
stabilizes \cite{Fonseca03} the ground state,
and the power $2.5$ comes from the putative scaling theory
\cite{Hamer90}
for the $(2+1)$-dimensional $q=3$ Potts model.
Note that the first-order phase transition 
also
obeys the remedied scaling theory
\cite{Fisher82,Challa86,Campostrini15,Campostrini14}.

The rest of this paper is organized as follows.
In the next section,
we present the numerical results.
The simulation algorithm is presented as well.
In Sec. \ref{section3},
we address the summary and discussions.

\section{\label{section2}
Numerical results}

In this section,
we present the numerical results
for the Potts model (\ref{Hamiltonian}).
To begin with,
we explain the simulation algorithm.

\subsection{\label{section2_1}Numerical algorithm}

We employed the exact diagonalization
method to simulate the Potts model (\ref{Hamiltonian})
for a rectangular cluster with $N \le 22$ spins.
In order to treat a variety of $N=16,18,\dots$ systematically,
we implemented the screw-boundary condition \cite{Novotny90}.
According to Ref. \cite{Novotny90},
an alignment of spins 
$\{ l_i \}$ with the first- and 
$\sqrt{N}$-th-neighbor interactions 
reduces to a rectangular cluster
under the screw-boundary condition.
Based on this idea,
we express 
the Hamiltonian as
\begin{equation}
\label{screw_Hamiltonian}
{\cal H}=
H_D(1)
+H_D \left( \sqrt{N} \right)
-\lambda \sum_{i=1}^N (R^+_i +R^-_i )
-H \sum_{i=1}^N  \frac{3}{2} \cos \frac{2\pi L_i}{3}
.
\end{equation}
Here, the diagonal matrix $H_D(v)$
denotes the 
$v$-th-neighbor interaction for an alignment $\{ l_i \}$;
that is,
 the diagonal element is given by
$\langle \{ l_i \} |H_D(v) | \{ l_i \} \rangle=
\langle \{l_i\} | T P^v | \{l_i\}\rangle$
with the translation operator,
$P|\{l_i\}\rangle = |\{l_{i+1}\} \rangle$,
and the Potts interaction,
$\langle \{l_i \} |T|\{l'_i\}\rangle=- \sum_{i=1}^N \delta_{l_i,l'_i}$.
The above formulae are mathematically closed;
however, for an efficient simulation,
Eqs. (9) and (10) of Ref. \cite{Nishiyama08b}
may be of use.

We performed the numerical diagonalization for the Hamiltonian matrix 
(\ref{screw_Hamiltonian}) by means of 
 the Lanczos method.
The single-magnon mass gap $m_1$
is given by 
\begin{equation}
\label{first_magnon}
m_1=E_1-E_0  ,
\end{equation}
with the ground-state ($E_0$) and 
first-excited ($E_1$) energy levels
within the zero-momentum sector.
Because the $N$ spins form a rectangular cluster,
the linear dimension is given by
$L=\sqrt{N}$, which 
sets a foundermental length scale in the subsequent scaling analyses.


\subsection{\label{section2_2}
Single-magnon mass gap $m_1$}

In this section, 
we investigate the single-magnon mass gap 
$m_1$
(\ref{first_magnon})
with the scaling theory \cite{Hamer90,Campostrini15}.
The first-order 
phase transition obeys the properly remedied scaling theory.

\begin{figure}
\includegraphics[width=100mm]{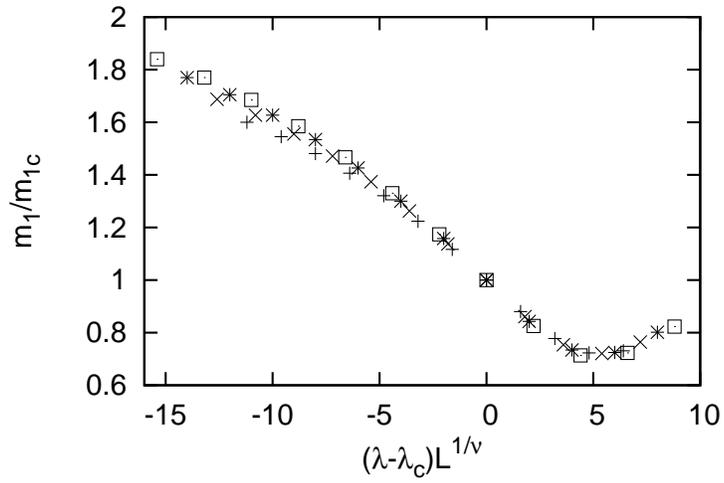}%
\caption{
\label{figure2}
The scaling plot, 
$(\lambda-\lambda_c)L^{1/\nu}$-$m_1/m_{1c}$,
is presented
for 
($+$) $N=16$,
($\times$) $18$,
($*$) $20$, and
($\Box$) $22$;
here, we adopted 
the scaling theory \cite{Hamer90,Campostrini15}
for the first-order phase transition.
The scaling parameters, 
$\lambda_c=0.8758$ and 
$\nu=0.5$,
are taken from the existing literatures, Refs. \cite{Hamer92} and \cite{Hamer90,Campostrini15}, respectively;
namely, there is no adjustable parameter involved
in
the  scaling analysis.
}
\end{figure}

In Fig. \ref{figure2},   
we present the scaling plot,
$(\lambda-\lambda_c)L^{1/\nu}$-$m_1 / m_{1c}$,
for 
($+$) $N=16$,
($\times$) $18$,
($*$) $20$,
and 
($\Box$) $22$.
Here, the symbol $m_{1c}$ denotes
$m_{1c}=m_1 |_{\lambda=\lambda_c}$ \cite{Campostrini15},
and the scaling parameters, $\lambda_c=0.8758$ and
$\nu=0.5$, are taken from Refs. \cite{Hamer92} and \cite{Hamer90,Campostrini15},
respectively.
We stress that there is no 
{\it ad hoc} fitting parameter
involved
in the scaling analysis.

The data in Fig. \ref{figure2}
seem to collapse into a scaling curve,
indicating that the simulation data already
enter the scaling regime.
Encouraged by this finding,
we turn to the analysis of the spectral properties.

\subsection{\label{section2_3}
Hierarchical spectral peaks $m_{2,3,\dots}$ via 
$\chi_{A}''(\omega)$}

Based on the finite-size scaling  \cite{Hamer90,Campostrini15}
demonstrated in the preceding section,
we analyze the dynamical susceptibility 
\begin{equation}
\label{dynamical_susceptibility_1}
\chi_A''(\omega)=
-\Im
\langle  0 | 
   A^\dagger 
(\omega +E_0 -{\cal H} +{\rm i} \eta)^{-1}
  A | 0 \rangle
,
\end{equation}
with
the ground-state energy (vector)
$E_0$ 
($|0\rangle$)
and
the energy-resolution parameter $\eta$.
Here, 
the perturbation operator
is set to
\begin{equation}
\label{perturbation_operator_1}
A= {\cal P} \left(
\sum_{i=1}^N J_i
\right)^2 ,
\end{equation}
with
\begin{equation}
J_i=
\left(
\begin{array}{ccc}
0 & {\rm i} & {\rm i} \\
-{\rm i} & 0 & 0 \\
-{\rm i} & 0 & 0 
\end{array}
     \right)
,
\end{equation}
and
the projection operator 
${\cal P}=1-|0\rangle\langle0|$.
We calculated the dynamical susceptibility
(\ref{dynamical_susceptibility_1})
with 
 the continued-fraction expansion \cite{Gagliano87}.
The dynamical susceptibility 
(spectral function) obeys
the
scaling formula
\begin{equation}
	\label{scaling_kasetu}
\chi_A'' \sim L^5 
f(\omega/m_1,(\lambda-\lambda_c)L^{1/\nu})
,
\end{equation}
with a certain scaling function $f$
\cite{Podolsky12,Nishiyama16}.

In Fig. \ref{figure4},
we present the scaling plot,
$\omega/m_1$-$L^{-5} \chi_A''(\omega)$,
with fixed $(\lambda-\lambda_c)L^{1/\nu}= - 4$
and $\eta=0.1 m_1$
for various 
$N=18$ (dotted),
$20$ (solid), and
$22$ (dashed);
here, the scaling parameters, 
$\lambda_c$ and $\nu$, are the same as those of 
Fig. \ref{figure2}.
The data appear to collapse into a scaling curve satisfactorily.

\begin{figure}
\includegraphics[width=100mm]{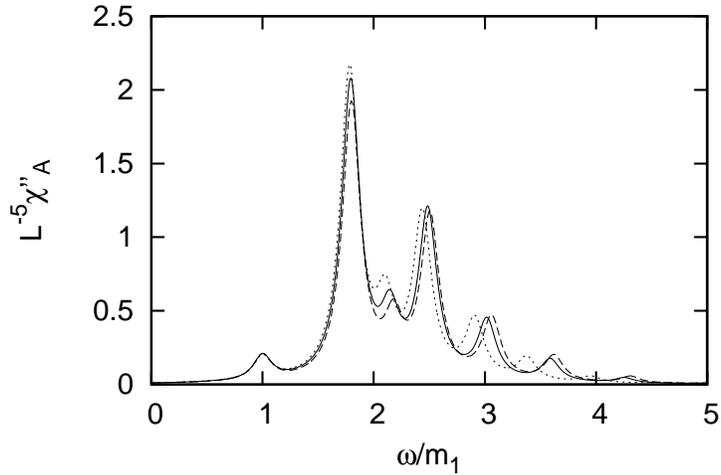}%
\caption{
\label{figure3}
The scaling plot,
$\omega / m_1$-$L^{-5}\chi_A''(\omega)$,
is presented with fixed $(\lambda-\lambda_c)L^{1/\nu}=-4$ and 
$\eta=0.1 m_1$
for various $N=18$ (dotted),
$20$ (solid),
and $22$ (dashed);
the scaling parameters,
$\lambda_c$ and $\nu$,
are the same as those of Fig. \ref{figure2}.
The signals, 
$\omega/m_1 \approx 1.8$,
$2.5$ and $3$,
correspond to the $m_{2,3,4}$ particles, respectively;
see text for details.
}
\end{figure}

Each signal in Fig. \ref{figure3}
is interpreted by the diagram
 in Fig. \ref{figure1}.
That is,
 the peaks around 
$\omega/m_1 \approx 1.8$,
$2.5$ and $3$ correspond to the $m_{2,3,4}$ excitations,
respectively.
The shoulder peak around $\omega / m_1 \approx 2$ should be
the two-magnon-spectrum threshold.
The signal $\omega/m_1\approx 3.5$ may be either the
$m_5$ particle or a composite one consisting of $m_1$ and $m_2$.
The ratios $m_{2,3,4}/m_1$ are estimated in the next section
more in detail.

Last,
we mention the choice of the perturbation operator
$A$ (\ref{perturbation_operator_1}).
In a preliminary stage,
we surveyed various types of the perturbation operators,
aiming to create the 
$m_{2,3,\dots}$ particles effectively.
Actually,
neither the first or the second term
of the Hamiltonian (\ref{Hamiltonian})
commutes with $A$;
otherwise, the susceptibility reduces to a mere
specific heat.
A key ingredient is that
 the exact diagonalization method permits us to treat
any off-diagonal operators.

\subsection{\label{section2_4}Universality of the scaled masses $m_{2,3,4}/m_1$}

In this section,
we devote ourselves to the analysis of 
the 
 scaled masses $m_{2,3,4}/m_1$;
 it is expected 
 that
 each ratio takes a constant value
in proximity to the transition point
 \cite{Hamer92}.

\begin{figure}
\includegraphics[width=100mm]{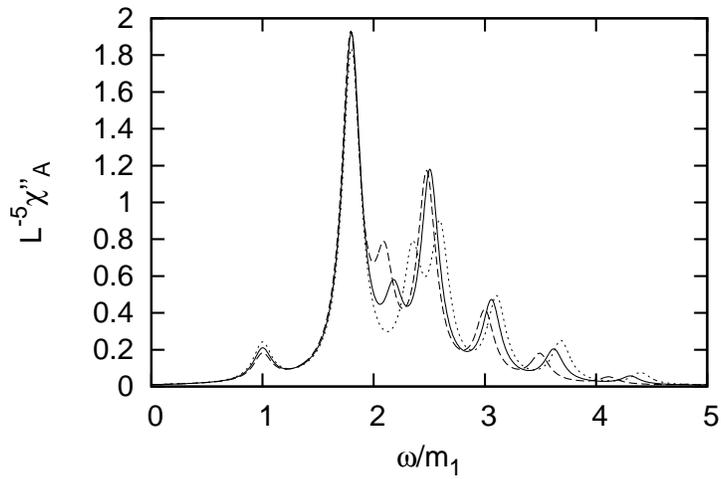}%
\caption{
\label{figure4}
The scaling plot,
$\omega / m_1$-$L^{-5} \chi_A''(\omega)$,
is presented with fixed $N=22$ and $\eta=0.1m_1$
for various values of the scaling parameter, 
$(\lambda-\lambda_c)L^{1/\nu}=-3$ (dotted),
$-4$ (solid), and
$-5$ (dashed);
the scaling parameters,
$\lambda_c$ and $\nu$,
are the same as those of Fig. \ref{figure2}.
Because the scaling parameter
$(\lambda-\lambda_c) L^{1/\nu}$ is raging, the curves do not necessarily
overlap.
The peak position seems to be invariant,
albeit with varying
the scaling parameter,
indicating that each scaled mass $m_{2,3,\dots}/m_1$
takes a constant value in proximity to the transition point.
}
\end{figure}

In Fig. \ref{figure4},
we present the scaling plot,
$\omega / m_1$-$L^{-5} \chi_A''(\omega)$,
 with fixed $N=22$ and 
$\eta=0.1m_1$
for various values of the scaling parameter,
 $(\lambda-\lambda_c)L^{1/\nu}=-3$ (dotted),
$-4$ (solid), and
$-5$ (dashes); the scaling parameters, $\lambda_c$ and $\nu$,
are the same as those of Fig. \ref{figure2}.
Note that these curves do not necessarily overlap, because 
the scaling parameter $(\lambda-\lambda_c)L^{1/\nu}$
is not a constant value; see Eq. (\ref{scaling_kasetu}) for the
scaling formula.
Each peak position seems to be kept invariant,
albeit with the scaling parameter varied.
As a result, we estimate
the scaled mass gaps as 
\begin{equation}
\label{masses}
(m_2/m_1,m_3/m_1,m_4/m_1)=
[1.80(3),2.5(1),3.05(25)] .
\end{equation}
Here, each error margin
was determined from the
 finite-size drift
 between $N=16$ and $20$;
a dominant source of the error margin comes from the oscillatory deviation
(an artifact due to the screw-boundary condition),
which depends on the condition 
whether the system size is a
quadratic number
$ N \sim 9,16,\dots $ or not 
$ N \sim 12( \approx 3.5^2),20(\approx 4.5^2),\dots$.

Each particle $m_{3,4}$ may possess a finite life time,
because it is embedded within the two-particle spectrum.
As a matter of fact, the data for 
$(\lambda-\lambda_c) L^{1/\nu}=-3$ (dotted)
in Fig. \ref{figure4}
exhibit split peaks around $\omega/m_1 \approx 2.5$,
indicating that the bound state $m_3$ has an appreciable
peak width,
$\Delta m_3 /m_1=0.3$.
Similarly, we observed 
$\Delta m_4 /m_1=0.35$ for the data with 
$N=18$ and 
$(\lambda-\lambda_c)L^{1/\nu}=-6$.
To summarize,
we estimate the intrinsic peak widths
(reciprocal life time)
as 
$\Delta m_3/m_1=0.3$ and 
$\Delta m_4/m_1=0.35$.
Each peak width is about one tenth of the corresponding
mass gap.

This is a good position to 
address an overview of the related studies.
First, 
for the classical
three-dimensional $q=3$ Potts model,
an estimate
$m_{2^+}/m_{0^+}=2.43(10)$ \cite{Falcone07} was reported;
the notation (symmetry index) is taken from the original paper.
This result may correspond to the present result, $m_3/m_1=2.5(1)$,
Eq.
(\ref{masses}),
supporting an ``approximate universality'' \cite{Hamer92} 
for the $q=3$ Potts model.
Second,
as for the Z$_2$ \cite{Agostini97}
and SU$(2)$ \cite{Fiore03}
gauge field theories,
the results,
$(m_{(0^+)'}/m_{0^+},
m_{(0^+)''}/m_{0^+},
m_{(2^+)'}/m_{0^+})=
[1.88(2),2.59(4),3.23(7)]$
and 
$(m_{(0^+)'}/m_{0^+},
m_{(0^+)''}/m_{0^+},
m_{2^+}/m_{0^+})=
[1.89(16),2.35(10),3.36(40)]$,
respectively, were obtained.
The hierarchical structures are quite reminiscent of
ours, Eq.  (\ref{masses}).
As a matter of fact,
for the gauge field theory,
the ``week $N$ dependence'' of the gauge group SU$(N)$ 
was reported \cite{Athenodorou15},
indicating a robustness of the hierarchy $m_{2,3,\dots}/m_1$.
For the SU$(3)$ gauge field \cite{Falcone07b},
a glueball mass, either 
$m_{2^+}/m_{0^+}=3.214(64)$ 
or $3.172(65)$,
was estimated.
The result may correspond to $m_4/m_1=3.05(25)$, Eq. (\ref{masses}).
Last, 
 for the Ising model
\cite{Caselle99,Lee01,Caselle02,Nishiyama08,
Dusuel10,Nishiyama14,Rose16,Nishiyama16},
estimates,
$m_2/m_1=1.82(2)$ \cite{Rose16}
and
$m_3/m_1=2.45(10)$ \cite{Caselle99}, were reported.
These results resemble to ours, Eq.  (\ref{masses}),
supporting a model-independence on the hierarchy.

Last, we address a remark.
Because 
the phase transition is discontinuous,
the continuum limit cannot be taken properly.
The above estimates such as the life time are specific to 
a lattice realization, although 
seemingly preferable scaling behavior was observed.
However, in an approximate sense,
the simulation results seem to be comparable with the 
related ones, 
as claimed by the preceeding studies \cite{Falcone07,Falcone07b,Hamer92}.

\subsection{\label{section2_5}
Continuum-threshold peak
via
$\chi_B ''(\omega)$}

As a comparison, we present a simulation result for 
$\chi_B''$ (\ref{dynamical_susceptibility_2}),
aiming to 
see
to what extent the 
 spectral weight is affected by the choice of the perturbation operator.
The dynamical susceptibility $\chi_B''$
is defined by
\begin{equation}
\label{dynamical_susceptibility_2}
\chi_B''(\omega)=
-\Im
\langle  0 | 
   B^\dagger 
(\omega +E_0 -{\cal H} +{\rm i} \eta)^{-1}
  B | 0 \rangle
,
\end{equation}
with a perturbation operator
\begin{equation}
B=
  {\cal P} \sum_{i=1}^N (R^+_i+R^-_i)
   .
\end{equation}
Note that the operator $B$ coincides with the
second term of the Hamiltonian (\ref{Hamiltonian}).
Hence, it 
exhibits the specific-heat-type singularity
$\chi_B'' \sim L^{2/\nu - 1}$ with $\nu=0.5$ \cite{Hamer90,Campostrini15}
right at the transition point.

\begin{figure}
\includegraphics[width=100mm]{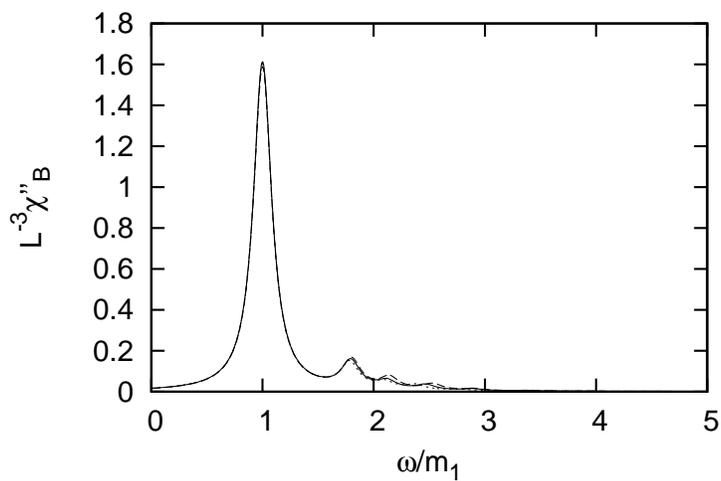}%
\caption{
\label{figure5}
The scaling plot,
$\omega/m_1$-$L^{-3} \chi_B''(\omega)$,
is presented with fixed 
$(\lambda-\lambda_c)L^{1/\nu}=-4$ and 
$\eta=0.1m_1$
for various 
$N=18$ (dotted),
$20$ (solid), and
$22$ (dashed);
the scaling parameters,
$\lambda_c$ and $\nu$, 
are the same as those of Fig. \ref{figure2}.
A faint signal around $\omega/m_1 \approx 2$
is attributed to the two-magnon-spectrum threshold.
The spectral weights for
$\chi_B''$
differ significantly from those for
$\chi''_A$.
}
\end{figure}

In Fig. \ref{figure5}, we
present the scaling plot,
$\omega /m_1$-$L^{-3}\chi_B''(\omega)$, 
with fixed 
$(\lambda-\lambda_c)L^{1/\nu}=-4$ and $\eta=0.1 m_1$ for 
various 
$N=18$ (dots), 
$20$ (solid),
and 
$22$ (dashed); the scaling parameters, 
$\lambda_c$ and $\nu$,
are the same as those of Fig. \ref{figure2}.
In contrast with $\chi_A''$ in Fig. \ref{figure3},
the susceptibility $\chi_B''$
detects the $m_{1,2}$ signals
and the two-magnon-spectrum-threshold  peak
$\omega/m_1\approx 2$; instead,
the bound-state hierarchy $m_{3,4}$ becomes hardly observable.

The result indicates that 
a naive external disturbance
such as the specific-heat-type perturbation $B$
does not create the bound states higher than $m_{1,2}$
very efficiently.
It is significant to set up the perturbation operator,
which does not commute with any terms of the Hamiltonian.
Note that the exact diagonalization method
allows us to survey various types of the
(off-diagonal) perturbation operators
so as to observe $m_{3,4,\dots}$ clearly.

\section{\label{section3}Summary and discussions}

The hierarchy
$m_{2,3,4}/m_1$ 
for the $(2+1)$-dimensional 
$q=3$ Potts model (\ref{Hamiltonian})
was investigated with the numerical diagonalization method.
The method allows us to calculate the dynamical susceptibilities
$\chi_{A,B}''$, Eqs. 
(\ref{dynamical_susceptibility_1}) and
(\ref{dynamical_susceptibility_2}),
via the continued-fraction expansion \cite{Gagliano87}.
Through the probe $\chi_A''$, we obtained
$(m_2/m_1,m_3/m_1,m_4/m_1)=[1.80(3),2.5(1),3.05(25)] $.
The particles $m_{3,4}$
acquire intrinsic
peak widths,
$\Delta m_3/m_1=0.3$
and $\Delta m_4/m_1=0.35$, respectively;
these spectra are embedded within the two-magnon spectrum.
According to Refs. 
\cite{Falcone07,Falcone07b},
the hierarchy $m_{2,3,4}/m_1$ of the Potts model
and that of
 the pure gauge theory are alike.
For instance, as for the Z$_2$-symmetric gauge group
 \cite{Agostini97},
there was reported
a hierarchy,
$(m_{(0^+)'}/m_{0^+},m_{(0^+)''}/m_{0^+},m_{(2^+)'}/m_{0^+})=
[1.88(2),2.59(4),3.23(7)]$, quite reminiscent of ours,
Eq. (\ref{masses}).

As a reference,
we calculated 
$\chi_B''$ (\ref{dynamical_susceptibility_2});
here,
the operator $B$ coincides with the second term of the
Hamiltonian
(\ref{Hamiltonian}),
and hence, it would be relevant to the experimental study.
It turned out that
the probe $\chi_B''$
is insensitive
to the hierarchy $m_{3,4,\dots}$,
indicating that the choice of the perturbation operator
is vital to observe $m_{3,4,\dots}$.
In this sense,
the exact diagonalization method
has an advantage in that 
we are able to treat
various 
perturbation 
operators so as to observe the hierarchy $m_{3,4,\dots}$ clearly.

\section*{Acknowledgment}
This work was supported by a Grant-in-Aid
for Scientific Research (C)
from Japan Society for the Promotion of Science
(Grant No. 25400402).

\end{document}